# Cooperative evolution of intraband and interband excitations for high harmonic generation in strained MoS$_2$

Meng-Xue Guan,[1,2] Chao Lian,[1] Shi-Qi Hu,[1,2] Hang Liu,[1,2] Sheng-Jie Zhang,[1,2] Jin Zhang,[1,2] and Sheng Meng[1,2,3, †]

[1] *Beijing National Laboratory for Condensed Matter Physics and Institute of Physics, Chinese Academy of Sciences, Beijing 100190, China*

[2] *School of Physical Sciences, University of Chinese Academy of Sciences, Beijing 100190, China*

[3] *Collaborative Innovation Center of Quantum Matter, Beijing 100190, China*

Modulating electronic structure of two-dimensional (2D) materials represents an exciting avenue for tailoring their optoelectronic properties. Here, we identify a strain-induced, cooperative effect of intraband and interband excitations contributing to high harmonic generation (HHG) in prototype dichalcogenide MoS$_2$ monolayer. We find that besides the dominant intraband contributions, interband current is also indispensable in modulating HHG. The HHG yields increase linearly with the compressive strain since flatter band dispersion and Berry curvature enhance both interband and intraband dynamics. Band structure can be retrieved with high reliability by monitoring the strain-induced evolution of HHG spectra, suggesting that strain not only provides an additional knob to control HHG in solids, but also marks a way towards a complete understanding of underlying microscopic mechanisms.

High harmonic generation (HHG) in solids is a new frontier in attosecond science and materials physics, which has been observed in various condensed-matter systems [1-8]. It not only provides an alternative route for generating brighter and more compact attosecond pulses [6], but also enables potential access to multi-petahertz electronics [9] and signal processing [10]. Meanwhile, HHG process stores information for reconstructing electronic band structure, affording new opportunities for studying strong field and ultrafast electron dynamics in condensed phases [11-14]. Due to their diverse electronic structures, the interactions of solids with an ultrafast, intense pulse cannot be simply described by the three-step model developed for atomic gases [15-19]. Instead, two distinct processes, i.e., intraband and interband transitions are believed to play decisive roles for solid-state HHG. The complex interplay between them can be controlled and utilized to improve the harmonic emission [19-21]. However, the relative contributions of intraband and interband processes and total excitations may vary in different materials and spectral ranges, leading to large differences among various experimental observations. For instance, Wang *et al*. demonstrated that the dominate generation mechanism in ZnO would switch from interband to intraband transitions as the driving wavelength increases [22].

In addition to optical modulation methods, such as tuning polarization direction of laser field [23] and ellipticity [24,25], solid targets provide possibilities to control HHG via tailoring their electronic structure, either by chemical or structural engineering [4,5,26,27]. It is not only a means to actively manipulate the generation process, but also an analytical tool to study the properties of a given system. In a variety of solid-state materials, the ability to continuously tune material properties is one of the most unique features of 2D crystals. Their electronic and optical properties are highly sensitive to external perturbations due to their atomic thickness [28]. In particular, monolayer transition-metal dichalcogenides (TMDs) such as $MoS_2$, have excellent mechanical properties that can withstand very large deformation before rupture, triggering a great interest in applying strain to tailor the crystal structure and modify their optoelectronic properties [29-32]. The effects of uniaxial compressive [33] or tensile strain [34] on the band structure of TMDs have been extensively studied by photoluminescence (PL) [35], Raman spectroscopy [36], and density functional theory (DFT) [37,38]. Thus, it is reasonable to expect that strain will serve as an effective tool to control HHG in TMDs that, to date, remains unexplored.

In this letter, using a full *ab initio* approach based on real-time time-dependent density functional theory (RT-TDDFT) [17,39-43], we demonstrate high sensitivity of HHG to atomic-scale structural deformation by applying uniaxial strain in monolayer $MoS_2$. To gain insight into electron dynamics underlying harmonic emission, we performed both semiclassical analysis and quantum mechanical dynamic simulations, showing that the HHG yield is sensitive and linearly dependent on strain, attributing to the fact that flatter band dispersion and Berry curvature help increase both intraband and interband contributions. The two contributions display an interesting cooperative effect with strain, with the predominant contribution by the intraband current. Meanwhile, we propose a way to reconstruct the target band structure by monitoring the strain-dependent HHG spectra. This work provides a major step towards tuning HHG characteristics of a solid target via mechanical strain engineering.

A single layer of $MoS_2$ (1L-$MoS_2$) is a direct gap semiconductor with an energy gap ($E_g$) of 1.8 eV, located at the valleys K and K' connected by time reversible symmetry [44]. Comparing with its bulk counterpart and, thanks to the enhanced electron-hole Coulomb interactions [45-47] and inversion symmetry breaking [48,49], 1L-$MoS_2$ exhibits improved HHG efficiency and the emergence of even harmonics, making it distinctive for HHG emission.

We consider both compressive and tensile uniaxial strain applied to the 1L-$MoS_2$ along, for example, the zigzag direction [Fig. 1(a)]. The results for other directions are qualitatively the same in our calculations (see Supplemental Material [50] for details). The strain is defined as $\varepsilon = \frac{a-a_0}{a_0}$, where $a_0$ and $a$ are the equilibrium and strained lattice constants, respectively. As a result, compressive strain will induce direct-to-indirect gap transition at $\varepsilon \approx -1\%$ and increase the band gap, whereas a tensile strain ($\varepsilon > 0$) only slightly reduces the energy gap [Fig.S1(c)], in good agreement with experiments and previous calculations [51-53]. To study the HHG response of strained 1L-MoS$_2$, we adopt a linearly polarized laser beam with time-dependent electric field $E(t) = E_0 cos(\omega t) exp[-(t-t_0)^2/2\sigma^2]$ [50]. The laser frequency and intensity are set as 0.32 eV and 2.24×10$^{10}$ W/cm$^2$, respectively.

To show the reliability of our theoretical approach, comparison between the calculated HHG spectrum and available experimental data [7] for strainless 1L-MoS$_2$ is shown in Fig. 1(b). The photon energy is set as one-sixth of the band gap of the primitive structure in both cases. Due to the low efficiency for higher-order responses, only the 6th to 13th harmonics are displayed in Ref. [7], where even-order emission is much weaker than the odd ones. The main features in experiment are well reproduced by our first-principles calculations, justifying the reliability of the present approach.

In order to investigate the role of structural deformation, HHG yield is monitored as a function of strain, shown in Fig. 2(a). The calculated spectra demonstrate the sensitivity of harmonic emission to small variations in the atomic structure. HHG yields for all orders increase by ~10% to ~150% when the lattice constant only changes by 3% under a compressive strain ($\varepsilon < 0$). However, for a tensile strain, enlarged strain reduces the HHG intensity at a lower rate. The above results indicate that the HHG yields can be continuously tuned by strain, and the compressive strain is more sensitive than tensile strain.

Figure. 2(b) shows that the relative change in HHG intensity ($\frac{\Delta I_\varepsilon}{I_0} = \frac{I_\varepsilon - I_0}{I_0}$, $I_\varepsilon$ and $I_0$ stand for the HHG yields with and without strain, respectively) is nearly linearly dependent on strain. However, the yields change at different rates on different HHG order. We performed analysis on the slope of each harmonic, as shown in Fig. 2(c). We notice that the slopes of even harmonics are generally larger than the odd ones, implying the different origins of the two kinds of harmonics. A semiclassical model based on acceleration theorem [54] was used to interpret the distinct polarization properties of odd and even harmonics in 1L-MoS2 [7]. The validity of the model requires the absence of the interband tunneling and scattering processes. In other words, the adiabatic approximation implies that intraband transitions make the dominate contribution to emission, where the carriers are accelerated within the single band driven by laser field.

To uncover underlying mechanism, we first start our discussion assuming that the semiclassical model (i.e. adiabatic evolution) is valid. Considering crystal symmetry restrictions, we expect that the emergence of even harmonics arise from materials' Berry curvature, while the odd harmonics exhibit a significant dependence on band dispersion [1,7]. In this scenario, both terms will change with uniaxial strain [Fig. S4]. For a more quantitative analysis, based on symmetry features in momentum space, Berry curvature and band dispersion can be represented by a series of spatial harmonics along K-Γ-K' direction:

$$\Omega(k) = \sum_{n=0}^{n_{max}} \Omega_n \sin(nkd) ,  \quad (1)$$

$$\varepsilon(k) = \sum_{n=0}^{n_{max}} \varepsilon_n \cos(nkd) . \quad (2)$$

Here, $d$ is the lattice constant, $\Omega_n$ and $\varepsilon_n$ are the $n^{th}$ Fourier coefficient of the Berry curvature and the lowest conduction band, respectively. In this semiclassical treatment, the dynamic evolution of the pre-existing wave-packet and the subsequently HHG emission are closely related to the magnitudes of $\Omega_n$ and $\varepsilon_n$ [2, 55].

Figures 3(a) and 3(b) show the dependence of $\Omega_n$ and $\varepsilon_n$ on strain. The standard deviation (σ) is used to quantify the amplitude of variation for two sets of data. A smaller $\sigma$ implies that the data points are closer to the mean of each set, making them decay faster as *n* increases [Fig. S5]. Considering that the wavepacket acceleration starts from the K and K' valleys of reciprocal space, the energy dispersion near K and K' should play the most crucial role in HHG emission and can be represented by the odd-order spatial harmonics [Fig. S6]. The decrease in the magnitude of $\sigma_\Omega$ and odd-order subset $\sigma_\varepsilon$ [brighter dots in Fig. 3(b)] under compressive strain suggests that flatter band dispersion and Berry curvature may help increase intraband contributions, leading to enhanced harmonic emission.

We emphasize that although the intuitive model using pure intraband dynamics can qualitatively capture the main features of HHG, it must be scrutinized when interband transitions also contribute to the generation process. For instance, in 1L-MoS$_2$, the role of Berry curvature in the intraband mechanism is to introduce an anomalous in-plane current that is perpendicular to the direction of the electric field of pump laser ($J \propto \dot{k} \times \Omega \propto E \times \Omega$), giving rise to only *perpendicular* component of even harmonics. Nevertheless, the oversimplified model fails to explain the experimental observation of even harmonics polarized *parallel* to the driving field (also see Fig. 4(a)), indicating that interband dynamics is indispensible.

For the interband process, the electron-hole pair is generated, then accelerated by the laser field and subsequently recombines coherently, emitting harmonic photons. Therein, electrons populated into the conduction bands $\Delta n_e(t)$ will oscillate in the momentum space, constituting Bloch oscillations together with intraband transitions. The value of $\Delta n_e(t)$ also oscillates in time. Therefore, accurate physical interpretation of HHG needs to consider both interband and intraband transitions.

In order to investigate the role of interband process, we compute the momentum resolved dynamics of excited electrons in conduction bands. Firstly, we have monitored the time evolution of the number of excited electrons by summing over all the conduction bands [50]. Figure 3(c) shows that more electrons are excited with lattice contraction during each half cycle, and then most of them return to ground state. The results indicate that under compressive strain, more carriers have been promoted to higher energy levels (interband) and to a broader momentum space (intraband), jointly contributing to harmonic generation.

The higher density of excited carriers by interband transitions would further flatten out the band dispersion and Berry curvature, thus increase the intraband contributions and lead a cooperative effect between them. The compressive strain will smooth overall band dispersion and Berry curvature distributions, therefore enhances intraband processes as well as interband excitations. Therefore, strain introduces cooperative intraband and interband evolutions in the HHG emission of 1L-MoS$_2$. We note that since accurate assignments of interband and intraband currents are gauge dependent [56,57], above analysis serves only as qualitative guide for comparison.

To verify our interpretation, momentum space resolved electron population is carried out by projecting time-evolved wavefunctions ($t = 60$ fs) onto the ground state Kohn-Sham orbitals. As shown in Figs. 3(d-f), it is clear that most of electrons are excited to the K and K' valleys of the Brillouin zone (BZ) of 1L-MoS$_2$, confirming our preceding assumption. Under compressive strain, electrons traverse a lager fraction of the BZ, the number of excited electrons at each momentum ($k_x$, $k_y$) increases simultaneously, which are direct evidences that interband and intraband dynamics cooperatively enhancing the harmonics yield. The results are reversed when tensile strain is applied.

To identify the relevant contributions of intraband and interband processes to the harmonic generation, we analyze polarization-dependent components of harmonics radiation, parallel and perpendicular to the linearly polarized fundamental field, as a function of strain [Figs. 4(a) and 4(b)]. As stated above, we can make a simple correspondence between intraband (interband) transitions and the perpendicular (parallel) component of even harmonics. Besides the obvious dominance of the intraband contribution ($\frac{I_{intra}}{I_{total}} > 65\%$), compressive strain enhances the proportion of interband contributions and the total HHG yield.

Therefore, the explicit roles of intraband and interband dynamics in the HHG radiation are identified. Each harmonic is produced predominated by the intraband current, while, interband contribution is essential in enhancing absolute HHG yield in the entire spectrum. The calculations show that the yields from both intraband and interband transitions increase, and the proportion of the interband contribution is simultaneously enhanced when external compressive strain is applied. Our results underpin the possibility to control the HHG generation process by band engineering via mechanical means, doping, or layer stacking. This conclusion is not limited to 1L-MoS$_2$, and applies for other TMD materials.

Apart from the above results, another interesting phenomenon captured from Fig. 2(c) is that the slopes exhibit a periodical variation. Moreover, the repeating energy interval corresponds exactly to the band gap of the equilibrium structure. For instance, the first order has the same slope as the seventh. Even if we change the frequency of the laser pulse, this phenomenon still prevails [Fig. S7]. Consequently, one could extract both the band gap and band dispersion (encoded in estimated effective mass of electrons) from the evolution of HHG in strained samples [Fig. S8].

This observation has recovered the known wisdom that interband dynamics plays a vital role in bridging below-band-gap and above-band-gap harmonics [17,22]. Beyond that, the result implies the possibility to retrieve electronic information of the solid target, such as the intrinsic band gap, by monitoring the strain-dependent HHG yield with a high confidence level [50]. In Ref. [11], Vampa *et al.* reconstructed momentum-dependent band gap of ZnO along the Γ-M direction by perturbing high harmonic generation with a weak second harmonic field. An external strain or additional field can both serve as a perturbative means. The advantage of the present approach is that the accuracy does not depend on the priori functional form of the trial band structure.

Strain-based reconstruction can be preceded by addressing more sophisticated effects. Figure 5 depicts the pattern evolution of some representative harmonics when the laser polarization is changed. It is clear that each harmonic exhibits a unique anisotropic response. Nonetheless, if the energy difference of two harmonics equals the band gap, they exhibit a similar pattern. For example, the 2nd and the 8th harmonic both resemble a dumbbell, while the 3rd and the 9th harmonic are similar to a spindle. In fact, each individual harmonic originates from different electronic wave packet trajectories in momentum space and, band structure information is imprinted on it [58]. For example, You *et al.* demonstrated that the anisotropic angular distribution of HHG provides direct insight into microscopic electronic dynamics in MgO [3]. Therefore, the differences between each pattern may uniquely tag the wave-packet trajectories and identify the band structure of the target material. However, detailed reconstruction methods warrant further exploration, which is beyond the scope of this work.

In conclusion, we predict that tailoring crystal structure via strain provides an effective tool to control HHG and helps clarifying the microscopic mechanism of HHG in solids. The HHG yield can be efficiently enhanced by applying compressive strain in 1L-$MoS_2$, beaconing a strong cooperative effect between intraband and interband contributions. Meanwhile, the relative contributions of the two are identified, i.e., intraband dynamics acts as the dominant factor of the HHG spectrum morphology, whereas interband dynamics is indispensible in modulating the absolute HHG emission. In addition, our results suggest the possibility to recover band structure based on solid-state HHG, where the intrinsic band gap is retrieved by monitoring the evolution in amplitude and orientation of HHG spectra. This work establishes a direct link between the HHG emission and underlying electronic dynamics, which would help design novel materials for tunable harmonic generation.

We acknowledge partial financial support from the National Key Research and Development Program of China (No. 2016YFA0300902 and 2015CB921001), National Natural Science Foundation of China (No. 11774396 and 11474328), and "Strategic Priority Research Program (B)" of Chinese Academy of Sciences (Grant No. XDB07030100). M. G. and C. L. contributed equally to this work.

[‡] smeng@iphy.ac.cn

# Figures

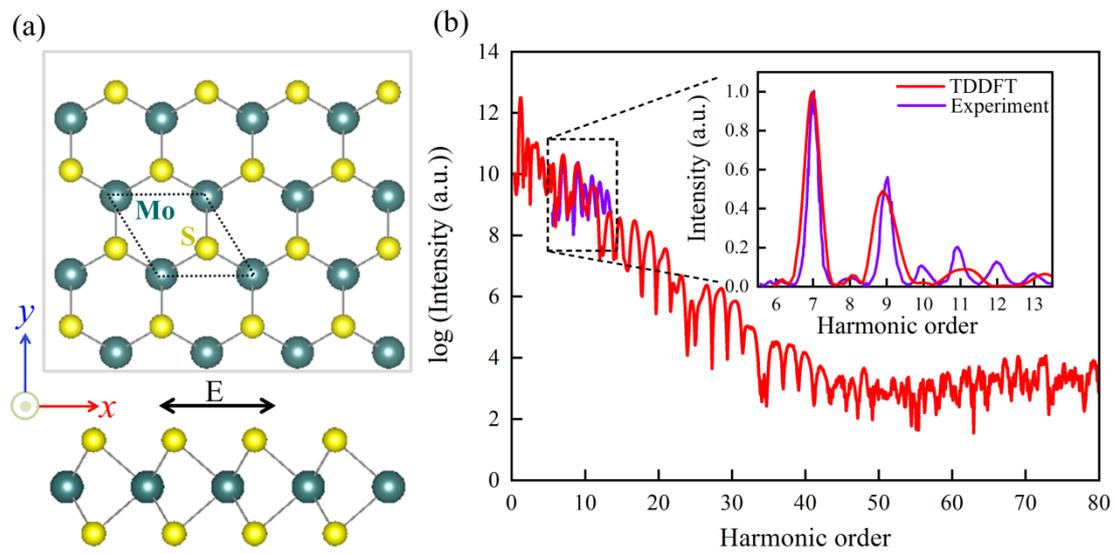

FIG. 1. (a) Schematic illustration of strained 1L-MoS$_2$ irradiated by linearly polarized laser pulse. The *x*, *y* axes are along the zigzag and armchair directions, respectively. (b) Computed HHG spectrum generated from primitive 1L-MoS$_2$ with the laser propagating along the zigzag direction (*x*). The inset shows the comparison between computed yields of the 6th to 13th harmonics (red) and the experimentally observed spectrum (purple) taken from Ref. [7].

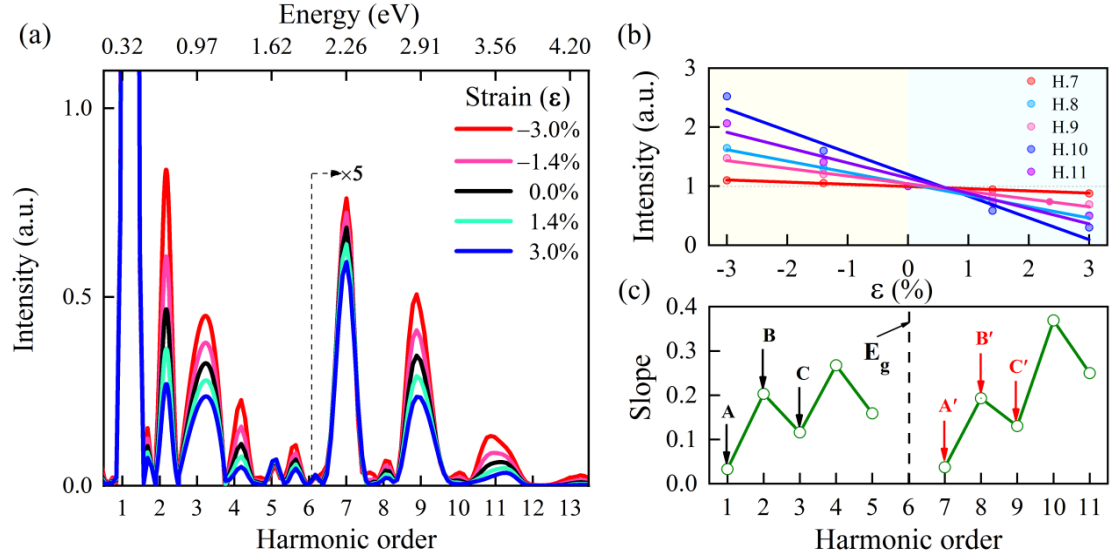

FIG. 2. Strain dependence of the yield for different harmonics of 1L-$MoS_2$. (a) Evolution of the normalized HHG spectrum under tensile and compressive strain. (b) The calculated harmonic yield as a function of strain for representative harmonics (colored dots), linear fit of the raw data can be used for all five harmonics (solid lines). Applying the same method to all harmonics and show their slopes in (c), the magnitude of the slopes change nearly periodically. Black and red arrows labeled as X and X' (X= A, B, …) denote the harmonics in the first and second cycle.

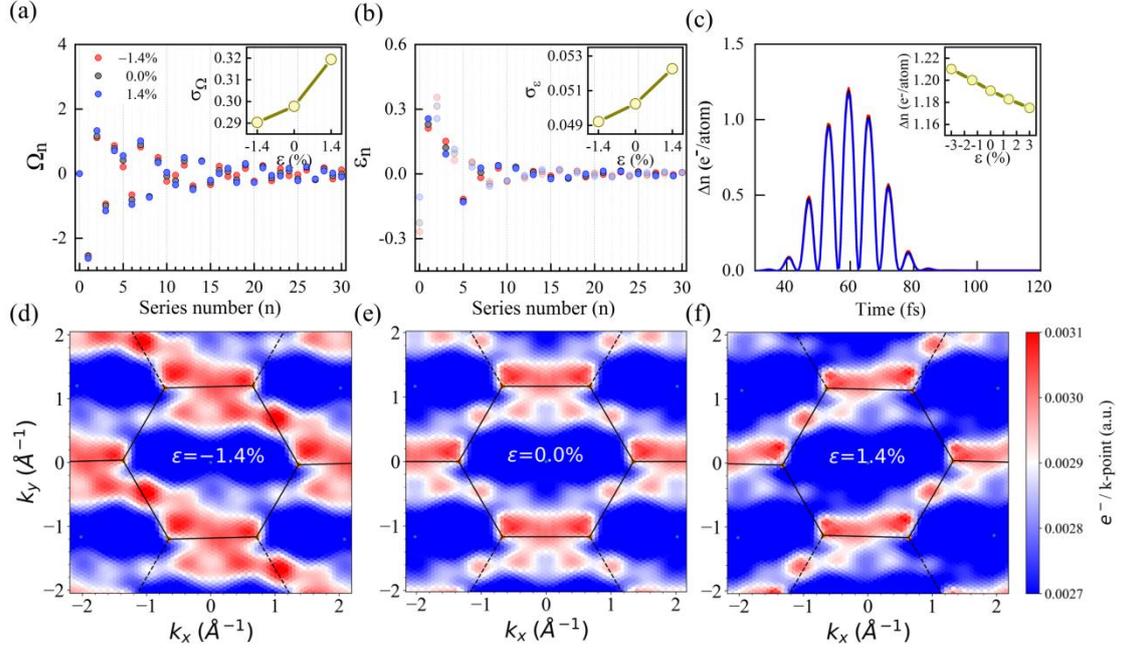

FIG. 3. Theoretical understanding of the strain modulation effect on HHG. (a, b) Fourier series coefficients of Berry curvature and the lowest conduction band dispersion with strain of −1.4%, 0, 1.4%. Standard deviations of data points are displayed in the inset. (c) Number of electrons excited to the conduction bands during the laser pulse. The inset shows the maximum value of excited electrons ($t = 60$ fs) as a function of strain. (d-f) False color representation of the momentum space resolved distribution of the excited electrons at $t = 60$ fs under strain −1.4% (d), 0.0% (e), 1.4% (f), respectively. The area bounded by black lines in each panel indicates the Brillouin zone.

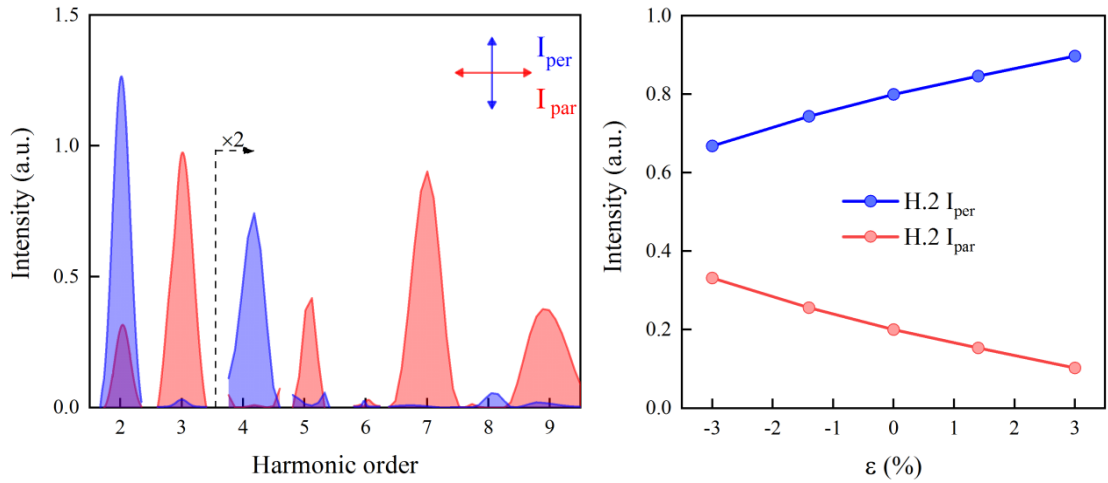

FIG. 4. Evolution of interband and intraband contributions to HHG with applied strain. (a) Computed harmonic yield from unstrained 1L-$MoS_2$ in the polarization basis perpendicular (blue) and parallel (red) to the linearly polarized excitation. (b) The perpendicular and parallel configurations of the 2nd harmonic (H.2) as a function of uniaxial strain.

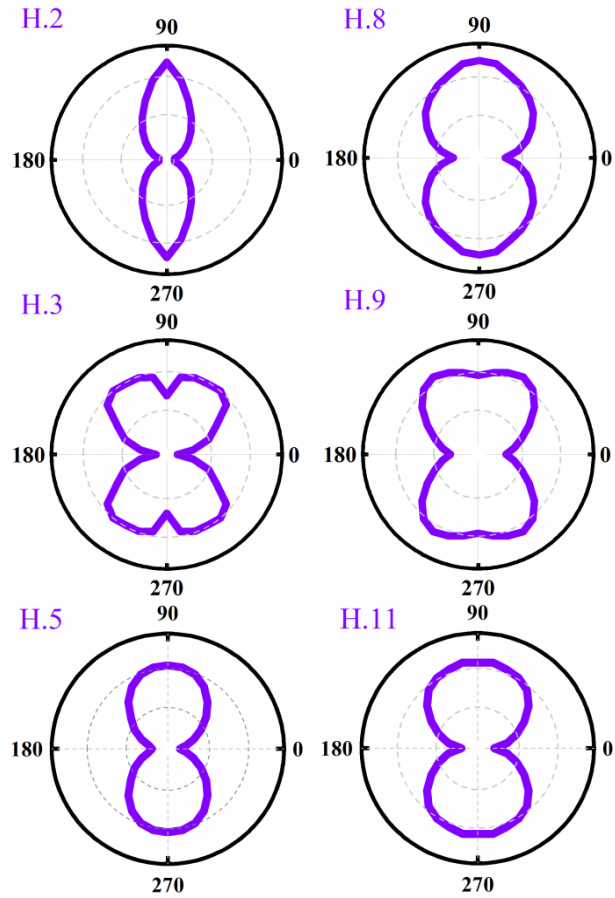

FIG. 5. Dependence of HHG on laser polarization direction of 1L-MoS$_2$. The fixed tensile strain (1.4%) is applied along the crystalline zigzag direction ($x$), while the laser polarization has a relative angle $\theta$ to it. The $\theta$ dependent harmonic intensity pattern for representative harmonics are shown. For simplicity, we use the intensity information of the primitive structure to renormalize all data. The numbers labeled on every subplot represent the value of relative angle.

Supplemental material for

# Cooperative evolution of intraband and interband excitations for high harmonic generation in strained MoS$_2$

Meng-Xue Guan,[1,2] Chao Lian,[1] Shi-Qi Hu,[1,2] Hang Liu,[1,2] Sheng-Jie Zhang,[1,2] Jin Zhang,[1,2] and Sheng Meng[1,2,3,†]

[1] *Beijing National Laboratory for Condensed Matter Physics and Institute of Physics, Chinese Academy of Sciences, Beijing 100190, China*

[2] *School of Physical Sciences, University of Chinese Academy of Sciences, Beijing 100190, China*

[3] *Collaborative Innovation Center of Quantum Matter, Beijing 100190, China*

## 1. Computational methods

The calculations are performed using the time dependent *ab initio* package (TDAP) as implemented in SIESTA, which has been successfully developed to describe strong field phenomena in solids [1-3]. Numerical atomic orbitals with double zeta polarization (DZP) are employed as the basis set. The Brillouin zone is sampled by $36 \times 36 \times 1$ Gamma centered $\boldsymbol{k}$-mesh with an energy cutoff of 250 Ry. The electron-nuclear interactions are described by Troullier-Martins pseudopotentials, in the adiabatic local density approximation (ALDA). It is important to note that within LDA, the band gap of semiconductor is underestimated and do not account for the excitonic effect. However, as the excitonic binding energy in TMDs is expected to be nearly independent of the uniaxial strain [4], thus relative variations in the electronic properties under strain and therefore HHG spectrum are credible.

To study the coherent interaction between a laser and strained 1L-MoS$_2$, we use a time step of 0.025 fs for quantum dynamic simulations lasting for 120 fs. The laser electric field $E(t)$ is described to be a Gaussian-envelope function,

$$E(t) = E_0 \cos(\omega t) \exp\left[-\frac{(t-t_0)^2}{2\sigma^2}\right]. \qquad (1)$$

Here, the width $\sigma$ is 12 fs, and $\omega$ is set to 0.32 eV (one-sixth of the band gap for equilibrium lattice) (see Fig. S1). The laser field value reaches the maximum strength $E_0 = 0.041$ V Å$^{-1}$ at time $t_0 = 60$ fs. For infinite periodic systems, we treat the field with a vector potential,

$$\vec{A}(t) = -c \int^{t} \vec{E}(t')dt'. \tag{2}$$

The time evolution of the wavefunctions are then computed by propagating the Kohn-Sham equations in atomic units (a.u.),

$$i\frac{\partial}{\partial t}\psi_i(\vec{r},t) = \left[\frac{1}{2m}\left(\vec{p} - \frac{e}{c}\vec{A}\right)^2 + V(\vec{r},t)\right]\psi_i(\vec{r},t), \tag{3}$$

where the vector potential $\vec{A}(t)$ appears in the kinetic term. Then time-dependent current can be obtained as,

$$J(t) = \frac{1}{2i}\int_{\Omega} d\vec{r} \sum_i \{\psi_i^*(\vec{r},t)\nabla\psi_i(\vec{r},t) - \psi_i(\vec{r},t)\nabla\psi_i^*(\vec{r},t)\}. \tag{4}$$

The HHG spectrum is obtained from time-dependent current through a Fourier transform,

$$HHG(\omega) = \left|\int_0^T \omega^2 J(t)\exp(-i\omega t)dt\right|^2. \tag{5}$$

The simulation of the dynamics of the excited electrons in momentum space were performed by projecting the time-evolved wavefuncitons ($|\psi_{n,\mathbf{k}}(t)\rangle$) on the basis of the ground-state wavefunctions ($|\varphi_{n',\mathbf{k}}\rangle$)

$$\Delta n_e(t) = \frac{1}{N_\mathbf{k}} \sum_{n,n'}^{CB} \sum_{\mathbf{k}}^{BZ} |\langle\psi_{n,\mathbf{k}}(t)|\varphi_{n',\mathbf{k}}\rangle|^2, \tag{6}$$

where $N_\mathbf{k}$ is the total number of the $\mathbf{k}$-point used to sample the BZ. The sum over the band indices $n$ and $n'$ run over all conduction bands. The momentum-resolved excited electron distribution, as shown in Figs. 3(d) to 3(f), is defined as:

$$\Delta n_e(\mathbf{k},t) = \frac{1}{N_\mathbf{k}} \sum_{n,n'}^{CB} |\langle\psi_{n,\mathbf{k}}(t)|\varphi_{n',\mathbf{k}}\rangle|^2. \tag{7}$$

## 2. Atomic structure of strained 1L-MoS$_2$ and its corresponding band structure

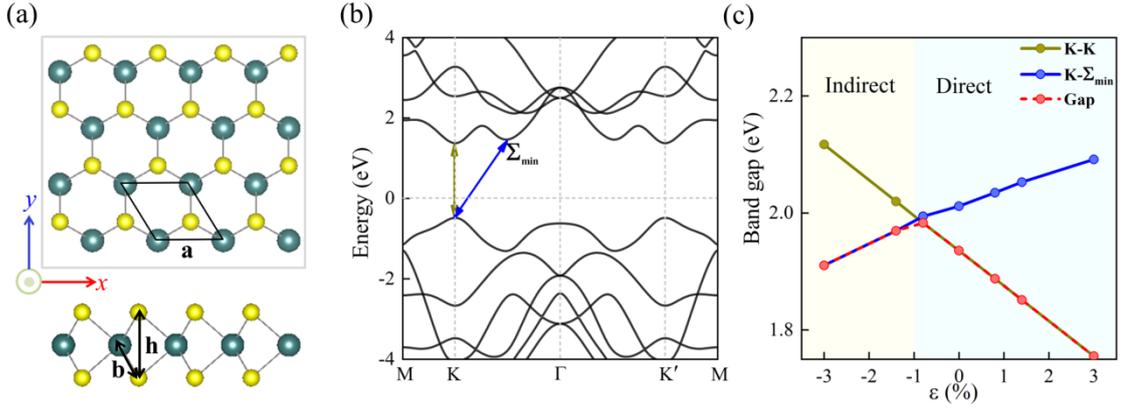

FIG. S1. (a) Atomic structure of 1L-MoS$_2$ with varied strain along the zigzag ($x$) direction. Detailed lattice parameters are listed in Table. S1. (b) Band structure of the primitive 1L-MoS$_2$, where the direct band gap locates at the two valleys K and K' of 1.94 eV. (c) The evolution of K-K and K-$\Sigma_{min}$ band gaps (as shown in (b)) with strain. The red dashed line gives the fundamental band gap.

Table S1. Calculated structural parameters and fundamental band gap for 1L-MoS$_2$ as a function of the applied strain

| $\varepsilon$ (%) | a (Å) (Mo-Mo) | b (Å) (Mo-S) | h (Å) | $E_g$ (eV) |
| --- | --- | --- | --- | --- |
| −1.4 | 3.079 | 2.377 | 3.128 | 2.02 (indirect) |
| 0.0 | 3.123 | 2.381 | 3.112 | 1.94 (direct) |
| 1.4 | 3.166 | 2.386 | 3.096 | 1.85 (direct) |

## 3. Applied Laser waveform and induced current

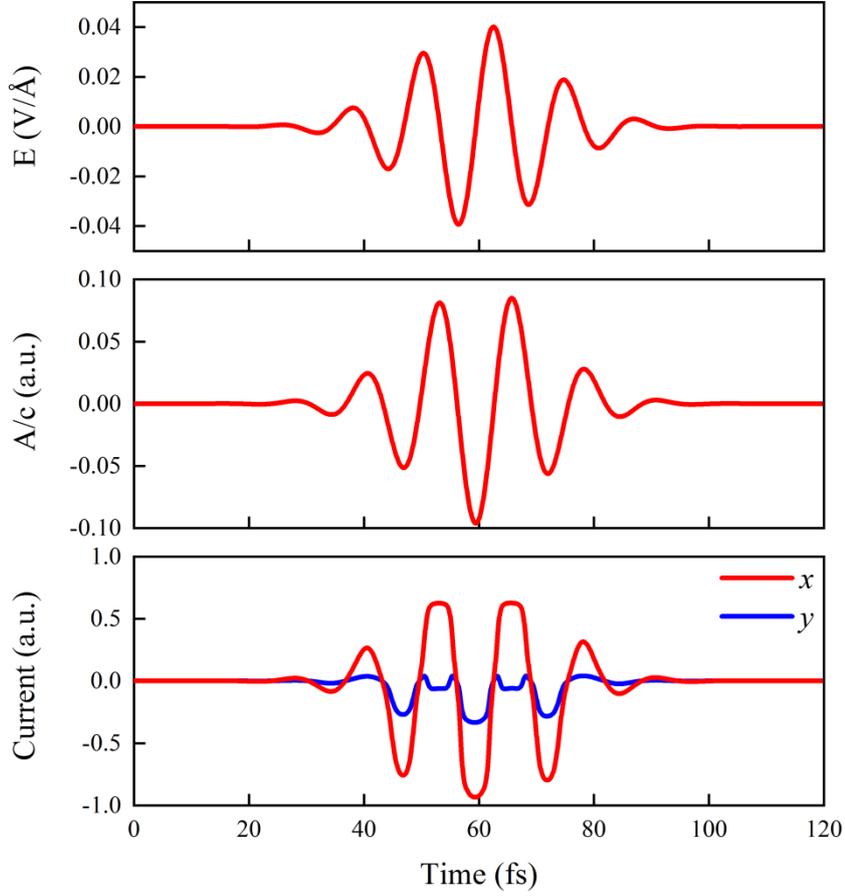

FIG. S2. Applied electric field (top), vector potential (middle) and induced electronic currents (bottom) that are parallel (red) or perpendicular (blue) to the incident polarization. The fundamental field centered at a photon energy of $\omega_0 = 0.32$ eV and polarization direction along the zigzag ($x$) direction. We sum the total current from different directions and then the HHG spectrum $I(\omega)$ is shown in Fig. 1(b).

## 4. Evolution of the normalized HHG spectrum with strain along the armchair direction

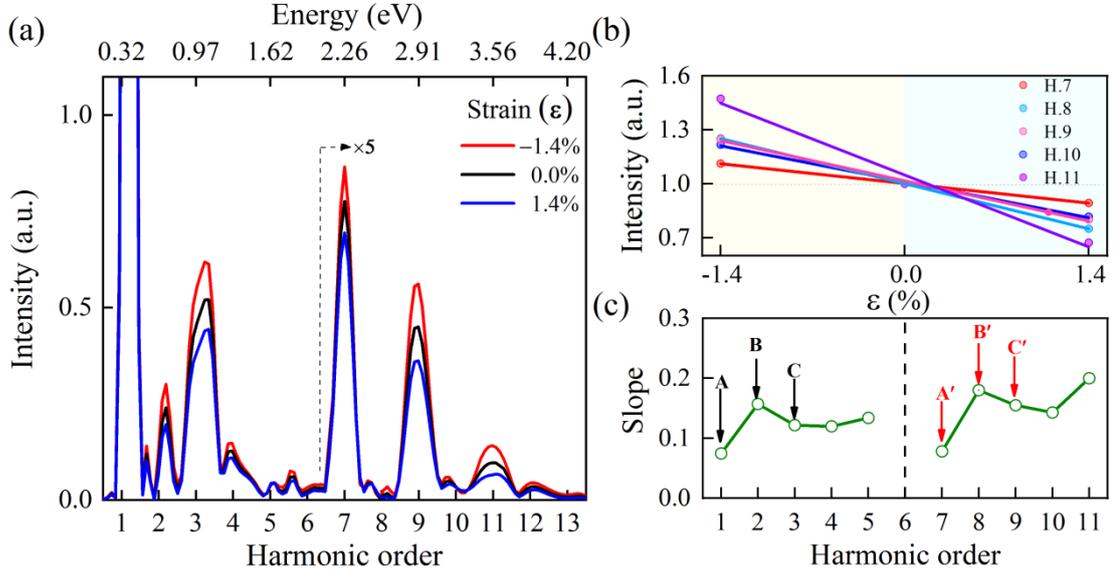

FIG. S3. Plots are analogous to Fig. 2. of main text, but correspond to the case when the uniaxial compressive and tensile strain are applied to the 1L-$MoS_2$ along the armchair direction.

## 5. Evolution of band dispersion and Berry curvature with strain

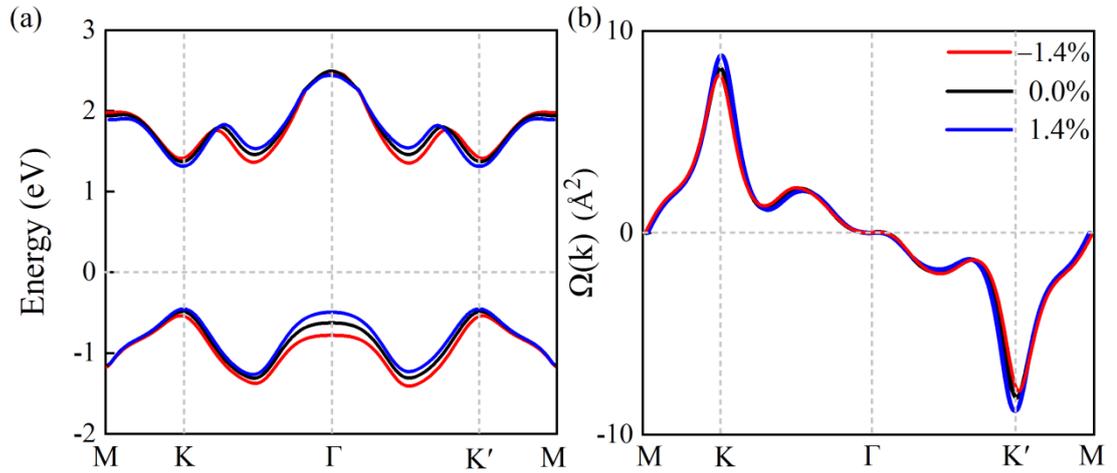

FIG. S4. (a) Evolution of band structure and (b) Berry curvature under three strain conditions, while the color of red, black and blue represent strain of $-1.4\%$, $0$, $1.4\%$, respectively.

## 6. Standard deviation of Fourier series coefficients and its relationship with the degree of flatness

The standard deviation $\sigma$ is defined as

$$\sigma = \sqrt{\frac{1}{n_{max}} \sum_{n=0}^{n_{max}} (x_n - \mu)^2}, \tag{8}$$

where $x_n$ are Fourier series coefficients of Berry curvature and the lowest conduction band dispersion, $\mu$ is the mean value of data points.

To show the relationship between the magnitude of $\sigma$ and the flatness degree, we compare two coefficient sets that have the same $\mu$ but different $\sigma$ [Fig. S5]. Similar with the functional form between Berry curvature and momentum, the curves can be expressed as

$$F(k) = \sum_{n=1}^{n_{max}} x_n \sin(nk). \tag{9}$$

It is clear that a smaller $\sigma$ indicates that the curve becomes flatter.

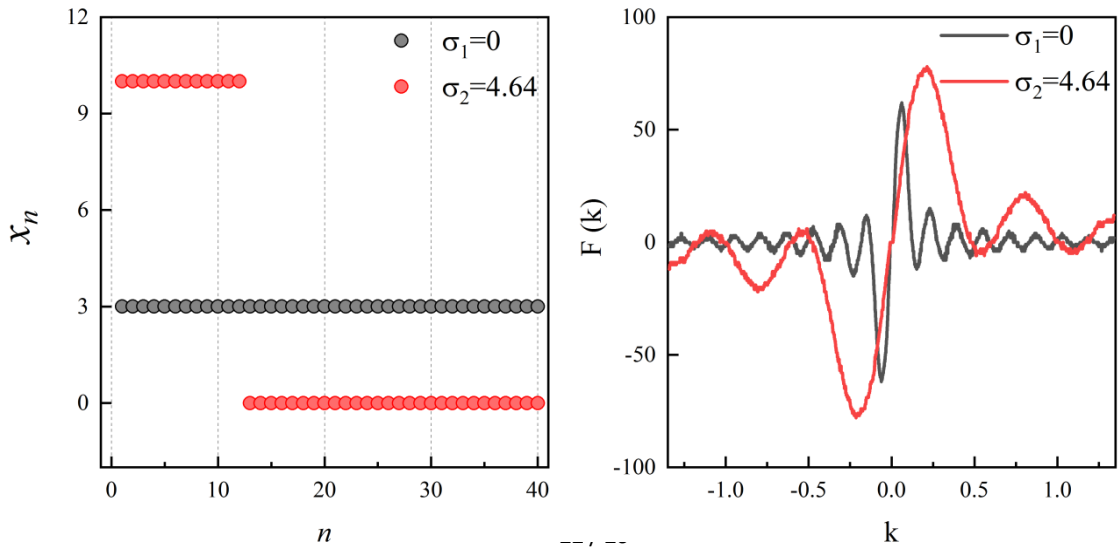

FIG. S5. (a) Two sets of coefficient $\{x_n\}$ with the same mean value $\mu = 3$, but have different standard deviation σ. (b) The dependence of the flatness degree on the magnitude of σ.

## 7. The contributions of odd and even-order spatial harmonics to the lowest conduction band

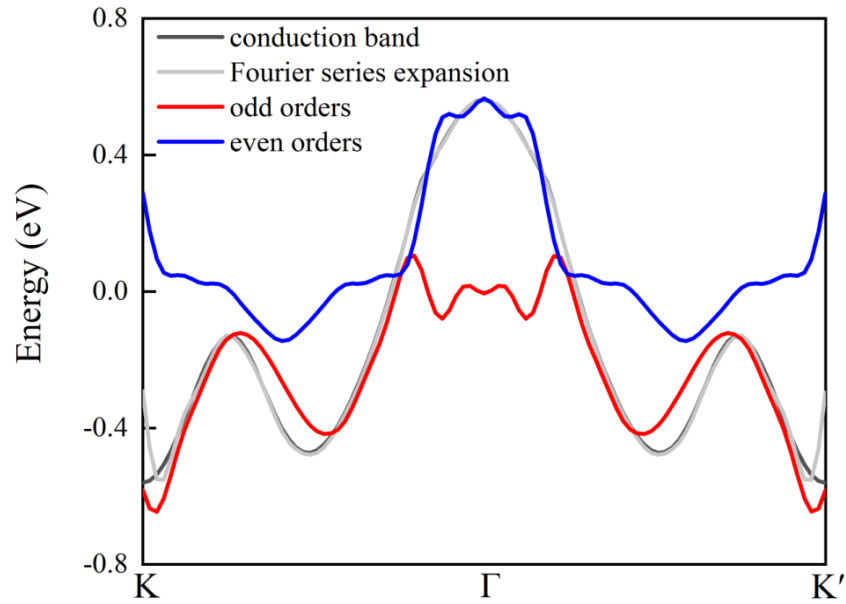

FIG. S6. Fitting of the lowest conduction band dispersion for strainless structure (black line) by Fourier series expansion that contains both odd and even-order spatial harmonics (grey line) and the relative contributions of either odd (red line) or even (blue line) order series. It is clear that around K and K', odd-order harmonics play the significant role in determining the HHG radiation. The results are similar under other strain conditions.

## 8. HHG spectra as a function of applied strain with photon energy equals to 0.24 eV

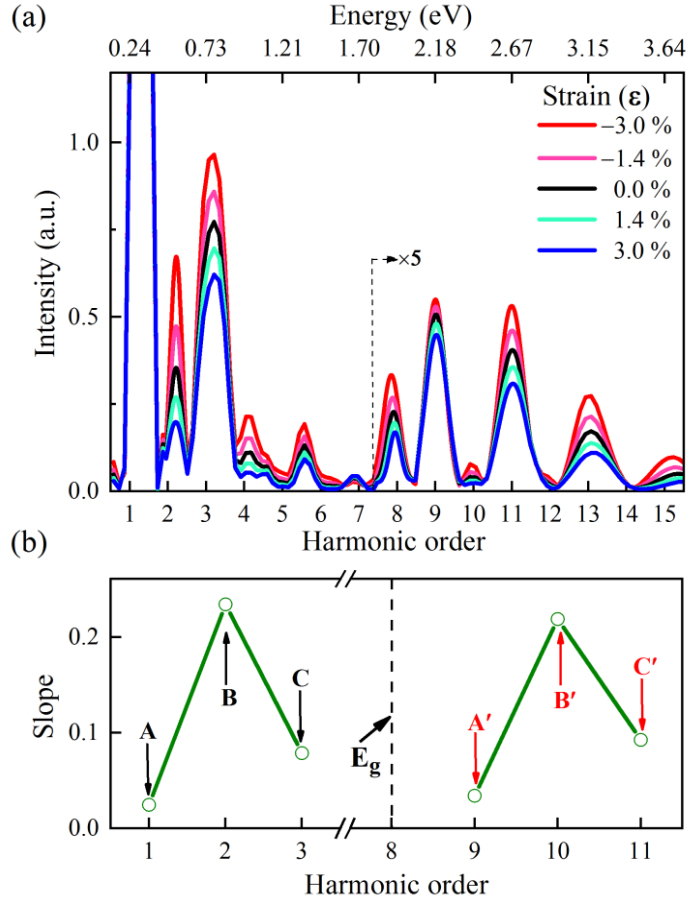

FIG. S7. (a) HHG spectra and (b) slope evolution with the fundamental field centered at a photon energy of 0.24 eV under strain along the zigzag (*x*) direction. The laser duration and intensity are same as those shown in Fig. S2.

## 9. Reconstruct effective masses by strain-induced band gap variation

The dependence of effective mass ($m^*$) on the second derivative of energy dispersion makes its value extremely sensitive to small variation in the band shape, making it critical a precise characterization of the electronic properties of the material in the DFT calculation. Here, we show how to estimate $m^*$ by strain-based reconstruction technique.

For strainless 1L-MoS$_2$, assuming that energy dispersion of the lowest conduction band and the highest valence band around K and K' are nearly parabolic, we then have,

$$\Delta E = \frac{\hbar^2(k_0 + \Delta k)^2}{2m} - \frac{\hbar^2 k_0^2}{2m}, \qquad (8)$$

where $k_0 = \left(0, \pm\frac{4\pi}{3a}\right) = (0, \pm 1.34)$ is the coordinate of K or K' in momentum space and $m = m_0 m^*$. For atomic units, if $\Delta k \approx 0$, then

$$\Delta E = \frac{k_0}{m^*} \Delta k = l * \Delta k. \qquad (9)$$

Here, $l = \frac{k_0}{m^*}$ is the slope of the energy change with displacement in the reciprocal space.

Figure S8 shows the evolution of the energy gap between the valence band maximum (VBM) and the conduction band minimum (CBM) at K with strain induced length variation, the corresponding slope equals to 3.14. Considering that both the shift of VBM and CBM contribute to the band gap evolution with nearly identical rates, the effective mass for K at VBM (or CBM) is estimated as $m^* = \frac{2k_0}{l} \approx 0.78$. On the other hand, $m^*$ were calculated by fitting the band structures at K along the K-Γ direction. In this way, the value obtained for hole at VBM, $m^* = 0.61$, and for electron at CBM, $m^* = 0.55$, fall well within the range of values already reported in the literature [5]. We note that, although the values of the effective mass calculated by the two methods are not strictly equal, strain-based reconstruction technique offers an effective way to obtain electronic structure information at a relatively reliable level.

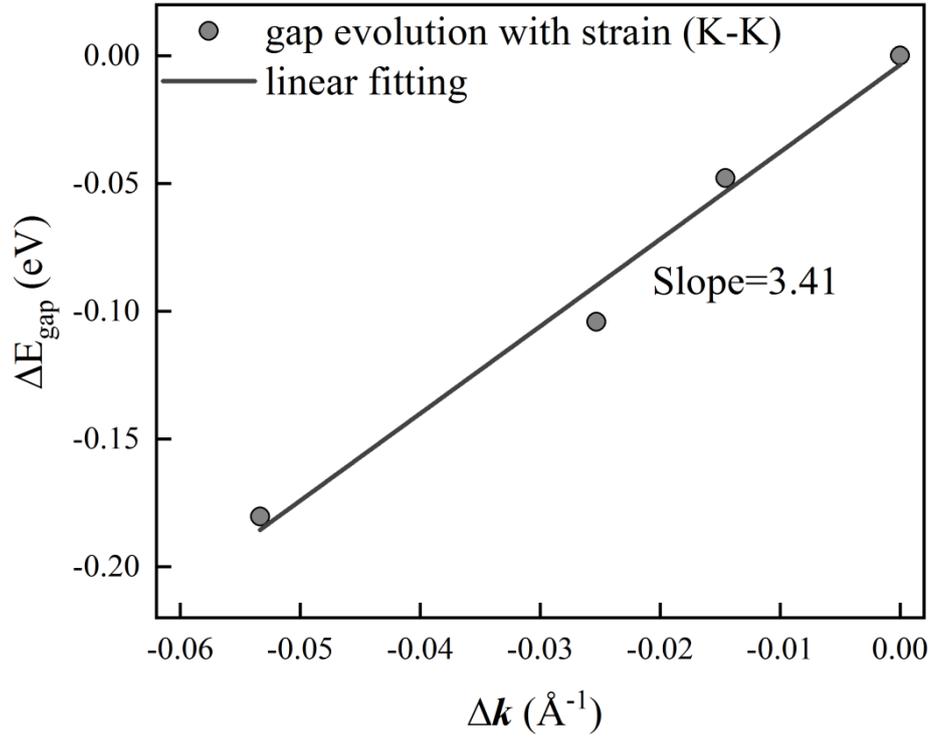

FIG. S8. Evolution of the band gap at K point with strain induced momentum change in reciprocal space.